\documentclass%[twocolumn]%
{jpsj3}

\title{On possible violation of the CHSH Bell inequality in a classical context}

\author{Bogus{\l}aw Broda\thanks{E-mail address: bobroda@uni.lodz.pl}\;and\;Micha{\l} Szanecki\thanks{E-mail address: michalszanecki@wp.pl}}
\inst{Department of Theoretical Physics, University of {\L}\'od\'z\\
Pomorska 149/153, PL--90--236 {\L}\'od\'z, Poland}

\abst{It has been shown that there is a small possibility
to experimentally violate the CHSH Bell inequality in a
``classical'' context. The probability of such a violation has
been estimated in the framework of a classical probabilistic model
in the language of a random-walk representation.}

\kword{Bell inequalities, CHSH inequality, hidden variables}

\begin{document}
\maketitle

The Bell inequality (BI) is supposed to discriminate between
``classical'' world and ``quantum'' one. Any classical theory,
i.e.\ causal theory governed by local (possibly, hidden)
variables, should fulfill the BI. Furthermore, any possible
violation of the BI is usually interpreted as a sign of
quantumness. In other words, it is commonly believed that only
quantum mechanics is allowed to violate the BI. Then, the
important question emerges: is it possible to violate the BI in
the framework of a classical theory? If so, the powerful role of
the BI could be a little diminished. It appears that the answer
can be in principle positive, but it depends on details. Actually,
the first such case is related to the notion of the postselection
\cite{Peres}, other one refers to the notion of the memory
loophole \cite{Barrett}, there is an analog of the detection
loophole presented in the NMR context\cite{Souza}, and finally
in optics\cite{Goldin}. One should emphasize that a lot of
them can be observed in laboratories.

In this letter, we will show, and this is our main aim, that in
finite number of measurement rounds it is possible to
``classically'' violate the BI, and we will estimate the
probability of such a violation in the framework of a simple
classical probabilistic model. To make things as simple as
possible, we will confine ourselves to the BI in the version of
Clauser, Horne, Shimony and Holt (CHSH) (see Shimony\cite{Shimony}, for a
contemporary introduction), and a simple classical Bernoulli-like
model.

We would like to strongly stress that our considerations have
nothing to do (though, perhaps, could be somehow linked to) with
real errors inevitably being encountered in actual measurements of
true experiments (see, our toy model as a simple demonstration of
the point). Otherwise, our conclusions would be entirely trivial.
In other words, our ``measurements'' are exact, but our model is
statistical.

%%%%%%%%%%%%%%%%%

Two (groups of) authors have already presented arguments similar to ours.
One should mention the two completely different, and even partially contradictory,
points of view: ``statistical'' one, advocated by Gill\cite{Gill1,Gill2},
and ``probabilistic'' one, pursued by Khrennikov and
others\cite{Avis,Khrennikov,Khrennikov1,Khrennikov2,Hess}.
As far as the statistical point of view is concerned, a notable
precedent of our work is a multi-thread and polemical study
undertaken by Gill\cite{Gill1,Gill2}. One of the relevant threads
of the Gill's papers is a discussion centered
around the role of statistics in violation of the CHSH Bell
inequality. Some authors (e.g.\ Accardi) propose classical, more or less controversial scenarios
which significantly, according to them, violate the CHSH Bell
inequality. Gill argues, using advanced statistical calculus, that
such large violations, suggested by them, are impossible.
Our approach, which is more elementary and explicit, bases on an example.
Besides, the goals (Gill's and ours) are different. Gill's goal is, in a sense,
``negative'', whereas our goal is ``positive'', or vice versa, it depends on the attitude.
It means that Gill finds a theoretical upper bound on the possible classical violation of the CHSH Bell 
inequality, whereas we explicitly show that actually the inequality can be classically violated.
In turn, the probabilistic point of view of Khrennikov and
others\cite{Avis,Khrennikov,Khrennikov1,Khrennikov2,Hess} emphasizes inadequateness of
the standard probabilistic framework for the description
of Bell type experiments. In fact, our critical, short
discussion of the standard proof of the CHSH Bell inequality given
around Eq.\eqref{CHSH Correlation Term Summation 1} and
\eqref{CHSH formulae 1} expresses our concern about the probabilistic
meaning of the inequality. Khrennikov {\it at al}\cite{Khrennikov2} even claim that in their
probabilistic formalism, they are able to support a model of Accardi,
which is supposed to significantly classically violate the CHSH
Bell inequality. But Gill proves\cite{Gill1,Gill2} that it is impossible,
and derives a bound (this is the partial contradiction mentioned in the beginning
of this paragraph). But we are not going as far as Khrennikov, and admit only a small,
not controversial possibility to violate the CHSH Bell inequality.

%%%%%%%%%%%%%%%%%

First of all, we would like to remove the seeming contradiction
between the well-known proof of the CHSH inequality and our a bit
controversially sounding statements. Namely, according to our
interpretation,
which plays an auxiliary role and is not crucial for our model,
one should note that a delicate point in the CHSH
argumentation consists in summing up (with one minus sign) and
collecting the all four terms
$a_{i}(\lambda)b_{j}(\lambda),\;i,j=1,2,$ i.e.\
\begin{equation}
a_{1}(\lambda)b_{1}(\lambda)-a_{1}(\lambda)b_{2}(\lambda)
+a_{2}(\lambda)b_{1}(\lambda)+a_{2}(\lambda)b_{2}(\lambda),
\label{CHSH Correlation Term Summation 1}
\end{equation}
under the common probability measure
$P(\lambda)\mathrm{d}\lambda$, yielding
\begin{equation}
 %\begin{split}
\int
P(\lambda)\left[a_{1}(\lambda)b_{1}(\lambda)-a_{1}(\lambda)b_{2}(\lambda)
+a_{2}(\lambda)b_{1}(\lambda)+a_{2}(\lambda)b_{2}(\lambda)\right]\mathrm{d}\lambda.
%\end{split}
\label{CHSH formulae 1}
\end{equation}
Here, as usual, $a_{i}(\lambda)=\pm1$, $b_{j}(\lambda)=\pm1$ are
values of the polarizations measured by the two (distant)
observers $\mathcal{A}$ and $\mathcal{B}$, respectively. The
indices $i,j=1,2$ label the two orthogonal directions of the
polarizers $A$ and $B$, and $\lambda$ is a collective hidden
variable governed by the probability distribution $P(\lambda)$
(see Haken and Wolf\cite{Haken}, for a very short introduction).
According to our interpretation the loophole in the CHSH
argumentation consists in the direct addition and integration in
\eqref{CHSH Correlation Term Summation 1} and \eqref{CHSH formulae
1}. One should take into account that, simply speaking, each of
the four correlations in \eqref{CHSH Correlation Term Summation 1}
is, in practice, being measured in different time instants. Since
hidden variables are in principle allowed to evolve in time, we
have in general different $\lambda$'s in each of the four terms,
and \eqref{CHSH Correlation Term Summation 1} and \eqref{CHSH
formulae 1} does not, in general, make sense.

If somebody is still skeptical concerning our above interpretation with the time $t$ in
its central role we offer the following additional remarks: 1)
Let $\lambda\equiv \lambda(t)=t$, for example. Does
eq.(\eqref{CHSH Correlation Term Summation 1}), eq.(\eqref{CHSH formulae
1}) or the standard normalization $\int P(\lambda)\mathrm{d}\lambda=1$ make sense?
Rather not. 2) The role of time is to break correlations, which
forbids making the valid sums of the type $a_{1}\pm a_{2}$ being
used in the proof of the CHSH inequality. Our mathematical
colleagues have already noticed some inconsistencies of this type,
referring it to the issue of different probabilistic spaces (see,
earlier references). 3) Our toy model presents a kind of direct
``experimental'' explanation of this phenomenon. 4) Finally, one
can skip this explanatory paragraph at all, if one dislikes it,
and directly jump to our model and the results which are
interpretation independent. Our interpretation plays only an
auxiliary role in our paper.

Obviously, the very fact that the proof has a loophole does not
automatically mean that the CHSH inequality can be actually
classically violated unless we show it explicitly. That will be
done in due course.

There is also another seeming contradiction in our proposal.
Namely, nobody is reporting an experimental classical violation of
the CHSH inequality due to finite statistics. The explanation is
very simple. Actually, the probability of such a violation in a
real experiment is very small. In fact, it depends on the number
of the measurement rounds. It will be shown that the greater the
number of the measurement rounds the lower probability of the
violation. The fact that the number of measurement rounds is
finite is crucial.

A conclusive explanation of the both above mentioned seeming
contradictions will be given in the form of the proposed model.
Mathematically, the model is formulated in probabilistic terms. In
other words, formally, speaking in the language of hidden
variables, one could state that the hidden variables are purely
random entities. From physical point of view, we could interpret
the randomness of the hidden variables in various ways. They could
be fundamentally random. Their randomness could follow from
complicated internal classical statistical mechanics. Or possibly,
their classical mechanical evolution could be fundamentally very
complicated, e.g.\ chaotic. Anyway, we are only interested in this
letter in a principal possibility of classical violation of the
CHSH inequality independently of a possible physical mechanism,
i.e.\ the nature of the hidden variables.

To quickly illustrate the point, first, we will present a toy
version of our model. To this end, let us confine ourselves (in
this toy version) to only four measurement rounds (in the full
version that number will be large). Let us perform the four
measurements in four different time instants, denoted: 1, 2, 3 and
4. We could even identify time with the hidden variable governing
the process, but it is optional. ``Accidentally'', it appears that
our experimental data are such as those given in Table~\ref{t1}.
Obviously, nothing forbids to obtain such data. As it is clear
from Table~\ref{t1} we deal with maximal possible violation of the CHSH
inequality. Namely, we obtain the (maximal) number 4 instead of
the classical bound given by the number 2 or the quantum bound
given by $2\sqrt{2}$. We can observe, enriching our earlier
arguments, that the constraints used in the standard proof of the
CHSH inequality do not work because the variables $a$ and $b$ in
different time instants (or for different hidden variables) are
totally unconstrained. Evidently, the data given in Table~\ref{t1}
constitute
\begin{table}
\caption{Data maximally violating the CHSH BI in
a classically allowed experiment.}
\label{t1}
\begin{center}
\begin{tabular}{|c|c|c|c|c|c|c|}
  \hline
 \textbf{measurement No.\ }&     &     &              &     &     &\textbf{contributions}  \\
 \textbf{(time instants)} & $\mathbf{a}$ & $\mathbf{b}$ & $\mathbf{c\equiv ab}$ & $\mathbf{i}$ & $\mathbf{j}$ &\textbf{to the CHSH BI} \\
 \hline
  1 & $+1$ & $+1$ & 1 & 1 & 1 & 1 \\
  2 & $+1$ & $-1$ & $-1$ & 1 & 2 & 1 \\
  3 & $+1$ & $+1$ & $1$ & 2 & 1 & 1 \\
  4 & $+1$ & $+1$ & $1$ & 2 & 2 & 1 \\
  \hline
  \multicolumn{6}{|r|}{\textbf{total CHSH correlation:}} &4\\
  \hline
\end{tabular}
\end{center}
\end{table}
a very particular configuration out of many others. How many? A
simple calculus, reduced for simplicity to $c\equiv ab$ (which is
equivalent to a more tedious calculus given in terms of $a$ and
$b$ separately) gives $2^{4}=16$ all configurations. We have
assumed that the $i$,$j$ pairs are fixed. This is a simplification
(also present in our full model)---in a more realistic situation
only the total number of the measurement rounds should be fixed
(to 4, in this example). Since there are 2 violating
configurations (one for 4 and one for $-4$), the total probability
$p$ of the violation of the CHSH inequality in our toy model
amounts to
\begin{align}
p=\frac{2}{2^{4}}=\frac{1}{8}=0.125.
\label{Violation probability
of the CHSH 1}
\end{align}

Now, let us consider the full model. In other words, we will
estimate the probability of the violation of the CHSH inequality,
in the spirit of the toy model, for a larger number of measurement
rounds. In terms of probability theory our model is described by a
Bernoulli process \cite{Feller}. But it is more convenient, from
intuitive point of view as well as from computational one, to
recast the model into (four-dimensional) random walks. As we have
just observed, we can think and work directly in terms of the
outcome $c\equiv ab$ instead of in terms of $a$ and $b$,
separately. In our random-walk representation the value $c=+1$
corresponds to the step forward, and the value $c=-1$ corresponds
to the step backward. The direction forward/backward $(\pm 1)$ is
``decided'' first. Next, the ``particle'' (we mean the abstract
particle of the random-walk representation) should be ``informed''
which dimension out of 4 should be followed. There are 4
dimensions corresponding to the 4 possibilities (pairs) given by
the orientations of the polarizers $A$ and $B$ ($i=1,2$ and
$j=1,2$). The total number of the steps in one of the four
dimensions is denoted by $n_{i}$ $(i=1,2,3,4)$. Since each step
assumes the value $+1$ or $-1$, the actual $i$th coordinate of the
position of the ``particle'' is the numerator of the $i$th
correlation. In other words, each of the 4 correlations entering
to the full CHSH correlation is of the form (``frequency
definition'')
\begin{align}
  \frac{\sum\limits_{\mbox{\tiny{1}}}^{n_{i}}\pm
  1}{n_{i}}\equiv\frac{m_{i}}{n_{i}}.
  \label{CHSH Correlation Term 2}
\end{align}
Therefore, the experimentally read CHSH correlation is given by
\begin{align}
  C=\frac{m_{1}}{n_{1}}-\frac{m_{2}}{n_{2}}+\frac{m_{3}}{n_{3}}+\frac{m_{4}}{n_{4}}.
  \label{CHSH Correlation 1}
\end{align}
Now, the inequality corresponding to the (rare) violating cases we
are interested in assumes the form
\begin{align}
  \left|C\right|>2.
  \label{CHSH Correlation 2}
\end{align}
For a large number of the measurement rounds, i.e.\
\begin{align}
   n_{i}\gg 1,
  \label{Emission number condition 1}
\end{align}
we can use a continuous approximation. The probability
distribution of the one-dimensional discrete random walk
\cite{Feller} is summarized by:
\begin{equation}
P_{n}(m)=\left\{
\begin{array}{ccl}
  \left(\frac{1}{2}\right)^{n}\frac{n!}{\frac{n+m}{2}!\;\frac{n-m}{2}!}, & \text{for} & m\equiv n(\rm{mod}\;2)\\
  \; & \;\\
 0, & \text{for} & m\not\equiv n(\rm{mod}\;2),
\end{array}
\right. \label{Discrete probability distribution 1}
\end{equation}
which in turn can be approximated, by virtue of the Stirling
formula, by
\begin{align}
  P_{n}(x)=\frac{1}{\sqrt{2\pi n}}\;e^{-\frac{x^{2}}{2n}}.
  \label{Continuous probability distribution 1}
\end{align}
Though our (auxiliary) random walk is four-dimensional, the
probability distribution for
\begin{figure}
\begin{center}
\includegraphics[scale=0.65]{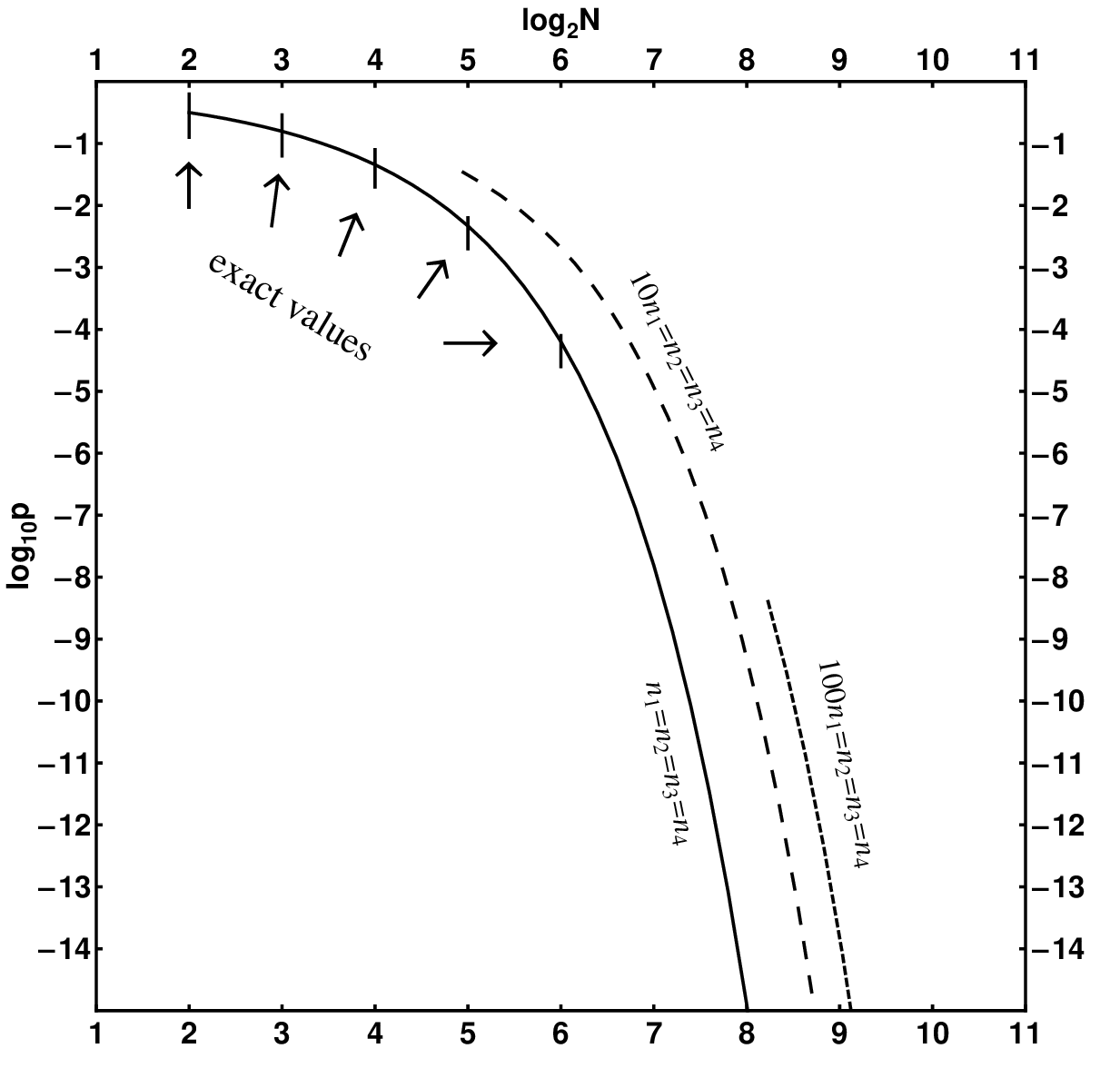}
\end{center}
\caption{Probability of the violation of the CHSH
BI versus the total number of the measurement rounds $N$
$(N=n_{1}+n_{2}+n_{3}+n_{4})$, in a doubly-logarithmic scale. 3
variants---solid line: $n_{1}=n_{2}=n_{3}=n_{4}$; dashed line:
$10n_{1}=n_{2}=n_{3}=n_{4}$; dotted line:
$100n_{1}=n_{2}=n_{3}=n_{4}$. The five vertical intervals
correspond to the exact (discrete) calculations easily performed
for small $N$ and equal $n_{i}$. The lowest points of the
intervals correspond to $C>2$, whereas the highest ones correspond
to $C\geqslant2$.}
\label{f1}
\end{figure}
each coordinate is independent. Therefore, the full
four-dimensional probability measure is given by
\begin{align}
  \mathrm{d}P_{n_{1}n_{2}n_{3}n_{4}}(x_{1},x_{2},x_{3},x_{4})=\frac{1}{\sqrt{2\pi n_{1}}}\frac{1}{\sqrt{2\pi n_{2}}}\frac{1}{\sqrt{2\pi n_{3}}}\frac{1}{\sqrt{2\pi n_{4}}}\;e^{-\frac{{x_{1}}^{2}}{2n_{1}}
  -\frac{{x_{2}}^{2}}{2n_{2}}-\frac{{x_{3}}^{2}}{2n_{3}}-\frac{{x_{4}}^{2}}{2n_{4}}}\;\mathrm{d}^{4}x.
  \label{Continuous probability distribution 2}
\end{align}
From mathematical point of view the task is to calculate the
probability of finding the walking particle in the
four-dimensional space outside the layer $L$ bounded by the two
hyperplanes $\mathcal{C}_{2}$ and $\mathcal{C}_{-2}$ (see, eq.\
\eqref{CHSH Correlation 1} and \eqref{CHSH Correlation 2}), where
\begin{align}
  \mathcal{C}_{\pm 2}:\;
  \frac{x_{1}}{n_{1}}-\frac{x_{2}}{n_{2}}+\frac{x_{3}}{n_{3}}+\frac{x_{4}}{n_{4}}=\pm
  2.
  \label{Hyperplanes condition 1}
\end{align}
For technical reasons we will ``isotropize'' the coordinates of
the probability measure by the following change of variables:
\begin{align}
  x_{i}=\sqrt{2n_{i}}\;z_{i},\;\;i=1,2,3,4.
   \label{Variables change 1}
\end{align}
Now, the (``isotropic'') probability measure \eqref{Continuous
probability distribution 2} is of the form
\begin{align}
  \mathrm{d}P_{n_{1}n_{2}n_{3}n_{4}}(z_{1},z_{2},z_{3},z_{4})=\frac{1}{\pi^{2}}\;e^{-{z_{1}}^{2}-{z_{2}}^{2}-{z_{3}}^{2}-{z_{4}}^{2}}\;\mathrm{d}^{4}z,
  \label{Continuous probability distribution 3}
\end{align}
whereas the equations of the bounding hyperplanes
\eqref{Hyperplanes condition 1} are
\begin{align}
\mathcal{C}_{\pm 2}:\;
\sqrt{\frac{2}{n_{1}}}\;z_{1}-\sqrt{\frac{2}{n_{2}}}\;z_{2}+\sqrt{\frac{2}{n_{3}}}\;z_{3}+\sqrt{\frac{2}{n_{4}}}\;z_{4}=\pm
2.
  \label{Hyperplanes condition 2}
\end{align}
The distance $d$ between the hyperplane given by the equation
$\sum\limits_{i}a_{i}x_{i}=b$ and the beginning of the coordinate
system is expressed by the formula:
\begin{align}
d=\frac{\left|b\right|}{\sqrt{\sum\limits_{i}{a_{i}}^2}}.
\label{Hyperplane distance}
\end{align}
Making use of the isotropy of the probability measure we can
rotate the hyperplanes so that they are orthogonal to the
coordinate $z_{1}$, say. Therefore, the probability $p$ we are
interested in is given by the integral
\begin{equation}
\begin{split}
p(n_{1},n_{2},n_{3},n_{4})=&2\frac{1}{\pi^2}\int\limits_{d}^\infty\mathrm{d}z_{1}\int\limits_{-\infty}^\infty\mathrm{d}^{3}z\;e^{-{z_{1}}^{2}-{z_{2}}^{2}-{z_{3}}^{2}-{z_{4}}^{2}}\\
=&\frac{2}{\sqrt{\pi}}\int\limits_{d}^\infty\mathrm{d}t\;e^{-t^{2}}\equiv\mathrm{erfc}(d)\\
=&\mathrm{erfc}\left(\sqrt{\frac{2}{\frac{1}{n_{1}}+\frac{1}{n_{2}}+\frac{1}{n_{3}}+\frac{1}{n_{4}}}}\right),
\end{split}
 \label{Probability formula p 1}
\end{equation}
where in the last line we have made use of \eqref{Hyperplanes
condition 2} and \eqref{Hyperplane distance}. Since $p$ is very
quickly damped by large $n_{i}$, the probability of the violation
of the CHSH inequality for $n_{i}\gg 1$ is very small.

A visual summary of our results is given in Fig.~\ref{f1}. Violation of
the CHSH BI, as well as its scale, is now evident. For simplicity,
in our model, we have assumed fixed values of $n_{i}$. In a more
realistic model only the sum $N=n_{1}+n_{2}+n_{3}+n_{4}$ should be
fixed, which would roughly correspond to a mean of
\eqref{Probability formula p 1} with respect to $n_{i}$. But,
obviously, our technical simplification, qualitatively, does not
change our final conclusion.

The continuous approximation is only a simplifying, technical
trick because discrete calculations, as combinatorial ones, are
straightforward only for small $N$. Nevertheless, we observe that
the continuous approach does work quite good also for small $N$.

Not to cause any misunderstandings in the field we would like to
stress that in the limit of \textit{infinite} number of
measurement rounds the CHSH Bell inequality is restored in the
classical domain. In other words, assuming the ``infinity''
condition as a premise in the CHSH Bell theorem removes the
classical violation but evidently the ``infinity'' condition is
experimentally unrealistic.

\begin{acknowledgments}

This work has been supported by the University of {\L}\'od\'z
grant and LFPPI network.

\end{acknowledgments}

\end{document}